\documentclass[twoside,11pt]{article}
\usepackage{amsmath,amsfonts,amssymb,verbatim,float,anysize,fancyvrb,graphicx,multicol,mathrsfs,mdframed}
\usepackage[utf8]{inputenc}
\usepackage[doublespacing]{setspace}
\usepackage[english]{babel}
\RequirePackage[colorlinks,citecolor=blue,urlcolor=blue]{hyperref}
\usepackage[usenames,dvipsnames, svgnames]{xcolor}
\usepackage{authblk}
\usepackage{multicol}
\usepackage{times}
\usepackage{subfig}
\usepackage{graphicx}
\usepackage{graphicx}
\graphicspath{{figures/}}
\usepackage{booktabs}
\usepackage[font=small,labelfont=bf]{caption}
\usepackage{amsfonts, amsmath, amsthm, amssymb}
\usepackage{wrapfig}
\usepackage[margin=2cm]{geometry}
\definecolor{mygreen}{rgb}{0,0.6,0.5}
\definecolor{myblue}{rgb}{0,0.45,0.7}
\definecolor{myred}{rgb}{0.8,0.4,0}
\definecolor{mygray}{rgb}{.6,.6,.6}

\parskip = \baselineskip
\usepackage{fancyhdr,amsmath,amsthm,amssymb,bm,url,enumerate,float} % Required for custom headers
\usepackage{lastpage} % Required to determine the last page for the footer
\usepackage{extramarks} % Required for headers and footers
\usepackage{graphicx,caption} % Required to insert images
\usepackage{lipsum} % Used for inserting dummy 'Lorem ipsum' text into the template
\usepackage{booktabs}
\usepackage[makeroom]{cancel}
    \usepackage[%
        backend=biber,
        style=numeric,
        natbib = true,
        backref=false,
        backrefstyle=all+,
        hyperref=true,
        sorting = none,
        maxbibnames = 10,
    ]{biblatex}
\newlength{\normalparindent}
\AtBeginDocument{\setlength{\normalparindent}{\parindent}}

\begin{document}
\setlength\parindent{24pt}

\title{Accounting for contact network uncertainty in epidemic inferences with Approximate Bayesian Computation}

\author{Maxwell H Wang\thanks{email: maxwang@hsph.harvard.edu} }
\author{Jukka-Pekka Onnela\thanks{email: onnela@hsph.harvard.edu}}
\affil{\small{Department of Biostatistics, Harvard University}}

\maketitle

\begin{abstract}
{When modeling the dynamics of infectious disease, the incorporation of contact network information allows for the capture of the non-randomness and heterogeneity of realistic contact patterns. Oftentimes, it is assumed that the underlying contact pattern is known with perfect certainty. However, in realistic settings, the observed data often serves as an imperfect proxy of the actual contact patterns in the population. Furthermore, the epidemic in the real world are often not fully observed; event times such as infection and recovery times may be missing. In order to conduct accurate inferences on parameters of contagion spread, it is crucial to incorporate these sources of uncertainty. In this paper, we propose the use of Mixture Density Network compressed ABC (MDN-ABC) to learn informative summary statistics for the available data. This method will allow for Bayesian inference on the epidemic parameters of a contagious process, while accounting for imperfect observations on the epidemic and the contact network. We will demonstrate the use of this method on simulated epidemics and networks, and extend this framework to analyze the spread of Tattoo Skin Disease (TSD) among bottlenose dolphins in Shark Bay, Australia.}
\end{abstract}

\section{Introduction}

In the study of emerging epidemics, it is important to understand the individual-level transmission dynamics of the contagion. Individual-based models can capture high resolution features of behavioral patterns, and can be useful for evaluating interventions that act on the individual level or predicting the spread of new contagions. To capture the dynamics of an epidemic, models typically attempt to describe the spread of disease via a set of mechanistic rules, which are typically modulated via a set of parameters that abstract the various aspects of disease transmission. For example, the per-contact transmissibility may capture the instantaneous rate of transmission across paths of potential transmission \citep{newman2002, wilson2008, groendyke2012}, and the expected duration of infectiousness captures the average time an individual remains infectious before recovery. Other parameters may capture the increase or decrease of viral load in an individual over time \citep{vrijens2005, jarvis2021}, the decrease in the probability of infection associated with protective mask-wearing \citep{bai2021}, or the proportion of individuals that are asymptomatic carriers of the contagion \citep{aguilar2020, luo2021}. Thus, an accurate disease model requires estimates of a range of parameters, as well as the quantification of the uncertainty associated with the estimates. Such estimation must often proceed from observed data, which is often subject to various sources of noise and missingness that are inherent to real-world observations. 

In this setting, it is natural to consider estimation and inference from a Bayesian perspective, which allows for the incorporation of prior information and serves as a principled method for accounting for various sources of uncertainty. In this paper, we will primarily consider two sources of uncertainty in data observed from epidemics: uncertainty regarding the contact structure of the population under study, and uncertainty regarding the contagion event times (e.g. times of infection).

The contact structure of a population is naturally captured by a contact network, which represents individuals as nodes and potential transmission paths as edges. Such networks have been used to model HIV \citep{kretzschmar1998, sloot2008, vieira2010}, measles \citep{groendyke2012}, COVID-19 \citep{liu2020, hambridge2021, tetteh2021}, and non-biological contagions, and contact networks are useful for identifying high-value targets for targeted interventions \citep{pastor2002, cohen2003}, informing methods to modify contact patterns to prevent contagion spread \citep{bu2013, youssef2013, zhang2022}, or as a predictor for the development of contagion \citep{harling2017, wang2023leveraging}.  While network data has traditionally been collected via surveys, advances in technology have allowed for the capture of location data via Bluetooth \citep{natarajan2007, eagle2009, sapiezynski2019} or wearable sensors \citep{barrat2008, cattuto2010, toth2015}. However, while much of the literature exploring the use of contact networks for epidemic models considers the networks to be fully observed, this is seldom the case. For example, social survey data may capture a reported relationship, but this perceived social link may only serve as a proxy for the true underlying contact type of interest, such as the duration of time two individuals spend within two meters of each other each week. Thus, observed contact data is often informative of the underlying patterns of contact, but the true contact network is typically not known with full certainty. 

Secondly, the full history of the epidemic under study is rarely fully observed. In particular, the times of infection and recovery for individuals may not be known, especially if the contagion of interest is asymptomatic, or if the status of individuals cannot be ascertained at all time points. Oftentimes, the disease status of individuals in the population may be known only at discrete points in time from diagnostic testing (such as for COVID-19) or observations for the presence of symptoms.

A common method to account for missing data is data-augmented Markov Chain Monte Carlo (MCMC). In these settings, unknown variables, which can include the unknown transmission trees within the contact network, the contact network itself, or missing event times, are treated as latent parameters that are jointly inferred upon along with parameters of interest. Data-augmented MCMC methods have been applied to network epidemics for a variety of applications \citep{britton2002, groendyke2011, groendyke2012, embar2014, bu2022}. However, when large amounts of data are missing, the latent variable space can become very high-dimensional. In such settings, data-augmented MCMC may not be computationally tractable unless the expressiveness of the model for either the epidemic or the network observations is appropriately limited. 

Approximate Bayesian Computation (ABC) is another subset of methods for Bayesian inference, first developed in the study of genetics \citep{tavare1997, beaumont2002}. Under the ABC paradigm, results are forward-simulated from the model, given a proposal value of the parameters, and these results are compared to the observed data. Proposal parameter values are accepted if the simulation results they produce are deemed to be similar enough to the observed data. In order to maintain computational tractability while also avoiding poor results from the curse of dimensionality \citep{blum2010}, the similarity of results is usually evaluated as the difference between two sets of summary statistics. ABC only yields exact inferences when the summary statistics are Bayes-sufficient and the acceptance threshold is zero; this is rare in practice, as finite-dimensional sufficient summary statistics are available only in the exponential family. However, approximate methods are nonetheless useful, as they allow for Bayesian inferences on sophisticated models that may not be analytically tractable. Thus, ABC is especially useful when considering models that are simple to describe in terms of mechanistic rules, but analytically complex.

In this paper, we will focus on the use of the Mixture Density Network-compressed ABC (MDN-ABC) \citep{hoffmann2022}. This method has been used for epidemic inferences on complex models when the contact network is fully observed \citep{wang2023}, but can be expanded to scenarios where the contact network is an additional parameter that must also be inferred. Unlike many other ABC approaches that depend on summary statistics evaluated from the trajectory of the epidemic, our method circumvents the need for summary statistic selection by employing a mixture density network to learn informative summary statistics. By augmenting the MDN-ABC procedure with an additional network sampling step introduced in \citep{young2020}, our method allows for sampling from the joint posterior distribution of the unknown network and the contagion parameters of interest. Previous methods that considered similar problems tend to focus on specific models for the contact network, such as simple statistical models \citep{britton2002, bu2022} or spatial models \citep{almutiry2020}. Contact network-augmented MDN-ABC (NA-MDN-ABC) provides a more generalized framework that allows for a high degree of flexibility in user specification of both the model for contagion dynamics, as well as the model for observations on the contact network.

In Section 2 of this paper, we will describe the theoretical background of MDN-ABC and its augmentation with a network sampling step. In Section 3, we will demonstrate the use of this method to conduct inferences on partially observed epidemics on noisily observed networks using a simulated example. In Section 4, we will use the NA-MDN-ABC to investigate the dynamics of tattoo-skin disease (TSD) in a population of bottlenose dolphins in Shark Bay, Australia. 

\section{Methods}

\subsection{MDN-ABC}
Approximate Bayesian Computation (ABC) is a likelihood-free method first applied to problems in population genetics \citep{fu1997, tavare1997}. The general idea behind ABC is to sample from the posterior distribution by accepting proposed parameter values that lead to simulated outcomes deemed similar to the observed data. If we consider $S(Y)$ as a summary statistic computed from the observed data $Y$, and $\delta$ is a acceptance threshold that determines how ``close" a simulated dataset must be to the observed dataset for the proposed parameter value $\theta^*$ to be accepted, a typical rejection ABC follows a simple rejection sampling scheme:
\begin{enumerate}
  \item[1.] Generate candidate parameter value $\theta'$ from the prior $\pi(\theta)$.
  \item[2.] Using $\theta'$, obtain a simulated datset $Y'$ (often via a simulated model).
  \item[3.] Calculate discrepancy between simulated and observed data: $d(S(Y_{obs}), S(Y'))$.
  \item[4.]  If $d(S(Y_{obs}), S(Y')) \leq \delta$, accept $\theta'$ as a sample from the posterior. 
  \item[5.] Return to Step 1 until a predetermined number of samples are obtained.
\end{enumerate}

In many applications of ABC, the choice for the summary statistic $S(Y)$ is not obvious. It is common to adopt \emph{ad hoc} summary statistics that intuitively provide information on the parameters of interest, based on metrics calculated from the observed data. For example, for studying epidemics spreading on networks, Dutta \emph{et. al} use the network induced by affected nodes at each time-step as one statistic \citep{dutta2018}. Other research has provideded methods to select a best subset from a set of summary statistics \citep{joyce2008, nunes2010, raynal2019}. These types of methods produce readily interpretable results, as long as the original set of candidate statistics were interpretable to begin with. However, intuitively apparent summary statistics may not be available for imperfectly observed scenarios; for example, one possibility is to utilize the mean of the recovery times as one summary statistic \citep{almutiry2020}, but this metric could not be calculated if recovery times are unknown. Furthermore, it is not always intuitively clear what metrics are most informative for the parameters of interest. 

Other methods seek to transform the observed data into useful summary statistics. These include estimation of the posterior mean as a summary statistic \citep{prangle2014}, maximizing Fisher information \citep{charnock2018}, or information-theoretic approaches such as minimizing Kullback-Leibler divergence \citep{chan2018}, minimizing posterior entropy \citep{nunes2010}, or maximizing mutual information \citep{chen2020}. It has been recently shown that the information-theoretic approaches are equivalent or special cases of minimizing the Expected Posterior Entropy (EPE) \citep{hoffmann2022}.

In this paper, we will consider the MDN-compressed ABC proposed in \cite{hoffmann2022}, where it is shown that it is possible to learn informative summary statistics by minimizing the Monte Carlo estimate of the Expected Posterior Entropy: 
\begin{equation}
\hat{\mathcal{H}} = -m^{-1}\sum_{i=1}^m \mbox{log}f(\theta_i, s(Y_i)),
\end{equation}
where $f(\theta,t)$ is a conditional density estimator that approximates the posterior, $\theta_i$ and $Y_i$ are joint samples from $p(\theta,Y)$, and $m$ is the number of samples in the minibatch. This work is closely related to conditional density estimation, where a Mixture Density Network \citep{bishop1994} is used to learn a conditional posterior density estimation that minimizes the EPE \citep{papamakarios2016}. Conditional density estimation yields a parameterized mixture model that can approximate the posterior density arbitrarily closely, given unlimited computational power. These conditional density estimators may be sometimes preferable to ABC sampling; notably, the conditional density estimator is able to provide a parametric approximation to the posterior density, while ABC can only produce samples from an approximate posterior distribution. However, conditional density estimators rely on parametric assumptions about the posterior distribution and, as such, do not enjoy the same asymptotic guarantees as ABC. Furthermore, in Section 2.3, we will augment the MDN-ABC with an additional sampling step to account for the network observation model; in order to marginalize over the high-dimensional contact networks, it is more convenient to draw samples from the posterior, rather than approximate a parameterized mixture distribution. Thus, instead of utilizing the components of the conditional density estimator, we will extract a single layer of the mixture density network to utilize as the summary statistics for ABC. This approach, dubbed the MDN-ABC, combines MDN and ABC methods to learn informative summary statistics using the MDN, while using traditional ABC sampling. Further details for this method can be found in \citep{hoffmann2022, wang2023}. In the following sections, we will consider augmenting the MDN-ABC to account for uncertainty in contact network reporting.

\subsection{Notation}
In the remainder of the paper, we will make use of the following notations. $Y$ is the observed epidemic data. We will consider cases where $Y$ is a binary vector of positive and negative test results of disease status; however, it is straightforward to consider other types of outcomes, such as event times or viral loads. $\theta$ are the parameters of interest, which determine contagion dynamics. These include parameters such as the per-contact transmissibility, or the mean infection time. $A$ is the true contact network that the epidemic propagates on. $X$ is the observed network data, which is assumed to take the form of measurements between pairs of individuals. For example, $X_{ij}$ could represent the number of times individual $i$ and individual $j$ were seen to interact during the duration of the study. While $X$ is informative of $A$, it does not perfectly capture $A$. Lastly, $\phi$ represents the parameters determining the distribution of the observed network data $X$, conditional on the true network $A$. This will be discussed in detail in the following section.

\subsection{NA-MDN-ABC for Network Inferences}

While MDN-ABC can readily be applied to epidemics on known networks \citep{wang2023}, it is often necessary to consider the uncertainty introduced by imperfect observations on networks. The unknown contact network can be considered a latent parameter that can be jointly inferred on, alongside the contagion parameters of interest. However, though ABC can potentially be expanded to high-dimensional applications, it is usually only practical to apply ABC when the parameter space is relatively low-dimensional. We propose to  focus on inferences for the contagion parameters while considering the true network as a nuisance parameter that is marginalized over. This is most sensible if the purpose of estimation of disease parameters is to parameterize contagion models for the prediction of disease, especially if the model is meant to be generalizable over populations that exhibit varying contact structures but are affected by the same type of disease, such as in the study of outbreaks in two cities with differing population densities. 

Young et. al \citep{young2020} discuss a simple framework for drawing Bayesian inferences on networks when only a proxy for the true network of interest is observed. Similar ideas were also previously explored in \citep{butts2003} and \citep{newman2018}. This approach consists of two primary components. First, the``data model" captures the probability distribution of the observed data, conditional on the true network: $P(X|A, \phi)$. Second, the ``network model" captures the \emph{a priori} probability of any given network: $P(A)$. This approach is most computationally efficient when full dyadic independence can be assumed (i.e. observations on each dyad are independent of all other dyads, and the probability of any edge existing is independent of any other edge existing). Under this assumption, it is relatively easy to express the full likelihood of $P(X|A, \phi)$ as a product of likelihoods. The factorizability of the likelihood leads to relatively simple sampling of the joint posterior $P(A, \phi|X)$.

In our paper, we are primarily interested in the joint posterior of the contagion parameters $\theta$, but the true network $A$ and the network parameters $\phi$ must included to account for uncertainty in network reporting. Thus, the full posterior that must be sampled from is in the form of $P(\theta,A,\phi|X,Y)$, where $X$ is the observed dyad-level data for the network, and $Y$ is the observed node-level data for the epidemic. To extract the marginal posterior of $\theta$, one simply marginalizes over $A$ and $\phi$. In the ABC setting, it is relatively simple to conduct such marginalization; joint samples of $(\theta, A, \phi)$ are drawn from $P(\theta,A,\phi|X,Y)$, and we only keep the samples of $\theta$.    

\begin{figure}[H]
\begin{center}
\includegraphics[page=1, width=6in]{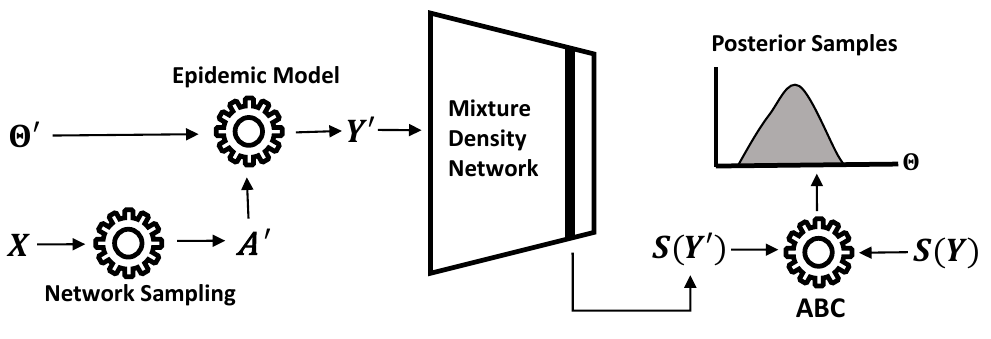}
\caption{Using the framework from \citep{young2020}, network samples $A'$ are drawn based on observed network data $X$. Proposal contagion parameter values $\theta'$ are drawn from the prior. $A'$ and $\theta'$ are passed through the epidemic model to generate a simulated output $Y'$. A summary statistic of $Y'$ is calculated via an MDN, yielding $S(Y')$. Finally, $S(Y')$ is compared to the summary statistics yielded by the original observed epidemic $Y$. If the acceptance condition is fulfilled, such that $d(S(Y'),S(Y)) \leq \delta$, $\theta'$ and $A'$ are accepted as part of the MDN-ABC posterior. To generate additional posterior samples, simply repeat the process with new values of $\theta'$ and $A'$.} \label{schematic}
\end{center}
\end{figure}

To proceed further, we make the following independence assumptions: 1) $X \perp Y |A$, 2) $\theta \perp \phi$, and 3) $\theta \perp A$. Here, Assumption 1 proposes that conditional on the true network, the observed network data $X$ is independent of the observed epidemic data. Essentially, if the true network is known, $X$ provides no additional information on $Y$, and vice versa. Note that this assumption would not hold true in situations where surveillance on the contact network is dependent on the disease status of individuals. For example, contact tracing, such as in the control of HIV, yields observational data on an underlying contact network. In this scenario, because the origin node of the contact trace is typically confirmed to be infected, the observations on the network would provide additional information on the spread of HIV beyond the information the true network itself carries. Assumptions 2 and 3 imply that in the absence of any observed information, the prior distribution of $\theta$ is independent of the nature of the true network, and how that network is observed. A common scenario where these assumptions may not hold is when individuals change their contact patterns based on disease status; for example, during the COVID-19 pandemic, some individuals likely self-isolated upon experiencing symptoms. However, Assumptions 2 and 3 are reasonable when 1) individuals do not change their behavior based on disease status or 2) network inferences focus specifically on the contact network \emph{prior} to the spread of contagion, and later behavioral changes are incorporated as dynamic evolutions of the pre-epidemic contact network. When these independence conditions are fulfilled, it is possible to consider the joint posterior in the simplified form: 
$$P(\theta, A, \phi|X,Y) \propto P(Y|A,\theta)P(A,\phi|X)P(\theta).$$
A derivation of this formula is found in the Appendix. This formulation lends itself to a straightforward MDN-ABC-based sampling scheme. A diagram is shown in Figure \ref{schematic}.

\begin{enumerate}
  \item[1.] Sample a candidate parameter value $\theta'$ from prior $\pi(\theta')$.
  \item[2.] Sample a candidate network $A'$ and network reporting parameters $\phi'$ from the joint posterior $P(A,\phi|X)$.
  \item[3.] Using $A'$ and $\theta'$, forward simulate to obtain a simulated epidemic dataset $Y'$.
  \item[3.] Given distance function $d$, compute distance between the summary statistics of the simulated and observed datasets: $d(S(Y_{obs}), S(Y'))$.
  \item[4.] Consider acceptance threshold $\delta$. If $d(S(Y_{obs}), S(Y')) \leq \delta$, accept $\theta'$ as a sample from the posterior. 
  \item[5.] Return to Step 1 until a predetermined number of samples are obtained.
\end{enumerate}

Note that due to the computational tractability of $P(A_t,\phi|X)$, there are a variety of ways to sample proposal networks for Step 2, including Gibbs sampling or MCMC. Similar to \citep{young2020}, we will implement this model in STAN, which utilizes a Hamiltonian Monte Carlo algorithm \citep{carpenter2017}. 

Note also that in order to perform this alogirthm, we first must obtain the summary statistics for $S(Y)$. Following \citep{hoffmann2022}, we will train the MDN by minimizing the Monte Carlo estimate of the EPE for $\theta$, the contagion parameters of interest. While one would ideally want the summary statistics to also be informative for $A$, the unknown contact network, it is not practical to optimize the EPE of such a high-dimensional object using an MDN, as the output size of the MDN would have to grow on the order of $n^2$. Thus, we will treat $A_t$ as a nuisance parameter, as our primary target of inference is $\theta$. This is sensible from a modeling standpoint, as a biological intuition of disease dynamics would be generalizable to other populations with differing contact networks. While the output of the NA-MDN-ABC is still the joint distribution $P(\theta,A,\phi|Y,X)$, we can simply marginalize over $A$ and $\phi$ to obtain the approximate posterior of $\theta$. For applications where inferences on $A$ are of greater importance and the epidemic process is used to obtain better estimates of contact network properties, consider the work by \citep{goyal2014}. Another potential direction is compression of network information, such as via the Graph Fourier Transform \citep{sandryhaila2013}. However, while methods such as the GFT are often able to express a network with high-fidelity using a low number of components, further work must be done to quantify the effects of training the MDN to learn summaries of the network, rather than the true network itself.

\section{Simulation Study}

In this section, we will consider a simulated SIR epidemic where the contact network and the event times are not observed. 

We assume that the simulated population consists of a set of $n$ individuals $N = {1,...,n}$, which are represented as nodes on a network $A$. The potential paths of transmission are represented by edges $E \subseteq N \times N$. We will consider two scenarios for degree distributions of the underlying network: Erdős–Rényi \citep{erdos1960} with mean degree 4, and log-normal with mean degree 4. The Erdős–Rényi network is generated as a random network, and the log-normal network is generated using the Chung-Lu model \citep{chung2002}, with the parameter $\sigma^2$ fixed at $0.5$. Both networks are initialized with mean degree 4.

In the compartmental SIR model, nodes can take on one of three states: ``susceptible" (S), ``infected" (I), or ``recovered" (R). Nodes begin in a susceptible state, with a subset of the population initiating in the infected state. Infected nodes cause neighboring susceptible nodes to progress to the infected state with a per-contact transmission rate $\beta$. Lastly, infected nodes progress to the recovered state with recovery rate $\gamma$. Nodes in the recovered state no longer infect neighboring nodes and are also unable to become re-infected themselves. More formally, following the definitions in \citep{fennell2016}, we can define $\beta$ and $\gamma$ as:
\begin{eqnarray}
\gamma & = & \lim_{\Delta t\to 0} \frac{P(X_{t+\Delta t}^i = R | X_t^i = I)}{\Delta t}. \\
\beta & = & \lim_{\Delta t\to 0} \frac{P(X_{t+\Delta t}^i = I \mbox{ via } j | X_t^i = S, X_t^j = I)}{\Delta t}.
\end{eqnarray}

To simulate uncertainty on epidemic event times, we will consider observations on the disease status. We periodically observe the binary disease status of each node: ``infected" for nodes in the susceptible or recovered state, and ``not infected" for nodes in the infected state. Thus, $Y$ takes the form of a vector of binary values.

Under this formulation, one can proceed with data-augmented MCMC sampling of the posterior distribution of $\beta$ and $\gamma$ when the contact network is known \citep{bu2022}. However, we will consider the case where the contact network $A$ is not directly observed. Instead, we will observe a proxy for the contact network, $X$. For our simulated observation model, we consider $X$ to be $N \times N$ matrix of dyadic encounters, such that $X_{ij}$ is the number of times nodes $i$ and node $j$ interact during the period of the study. We model $X_{ij}$ as a Poisson-distributed count with a rate dependent on whether or not the dyad in the true network $A$ is an edge or a non-edge, such that:
$$X_{ij}|A_{ij}=0 \sim \mbox{Poisson}(\lambda_0),$$
$$X_{ij}|A_{ij}=1 \sim \mbox{Poisson}(\lambda_1).$$

Furthermore, the \emph{a priori} probability of any edge existing is consdered to be a constant $\rho$:
$$A_{ij} \sim \mbox{Bernoulli}(\rho).$$

For our simulations, we will initiate the contagion parameters with values $\beta = 0.15$ and $\gamma = 0.1$, and network observation parameters $\lambda_0 = 1$ and $\lambda_1 = 8$. To begin the simulation, we randomly select 5 nodes as the origin of the infection. We continue our simulated epidemic for 50 timesteps, and obtain each node's status every 7 timesteps. The status of the node is considered to be ``1" if the node is infected, and ``0" if the node is susceptible or recovered. If each timestep is considered to be a day, this would correspond to a weekly testing cadence. We will generate a single instance of this simulation to serve as the ``true" epidemic -- our observations on the true epidemic will be $Y$, the vector of binary values of individual-level test results, and $X$, the count of interactions between each pair of individuals in the population. In a real study employing MDN-ABC (which will be explored in Section 4), this is analagous to the observed epidemic that takes place in reality. 

\subsection{MDN-ABC Settings}

The priors for both $\beta$ and $\gamma$ are set to $\mbox{Gamma}(2,4)$ distributions. The prior for $\lambda_0$ was set to $\mbox{Gamma}(2,4)$, while the prior for $\lambda_1$ is set to $\mbox{Gamma}(24,2)$. The prior for $\rho$ was set to $\mbox{Beta}(2,2)$. 

In order to train the MDN, we generated $5\times10^6$ samples for training, and $2.5\times10^6$ samples for validation. As described in Section 2, we drew contagion parameters $\theta$ from their priors. In order to sample contact networks for the simulations, we will utilized the HMC method described in \citep{young2020}. We trained our MDN for two gamma-distributed components. Our neural network was a fully-connected architecture, with 6 hidden layers. We extracted the last hidden layer, with 15 neurons, as the vector of summary statistics. We used an Adam optimizer with a learning rate of $5\times10^{-5}$. At each epoch, loss on the validation set was calculated. Once 10 epochs elapsed without an improvement in validation loss, training was terminated. Once the summary statistics were trained, we re-utilized the training samples in a simple rejection-ABC algorithm. We chose the discrepancy function to be the Euclidean distance. The summary statistic vector for each training sample was compared to the summary statistics yielded by the original ``true" epidemic. The 0.02\% of proposal parameters corresponding to the simulation results with the lowest discrepancy values were accepted as part of our MDN-ABC posterior. This corresponded to roughly 1000 posterior samples. 

\subsection{Results}

\begin{figure}[H]
\begin{center}
\includegraphics[page=1, width=5in]{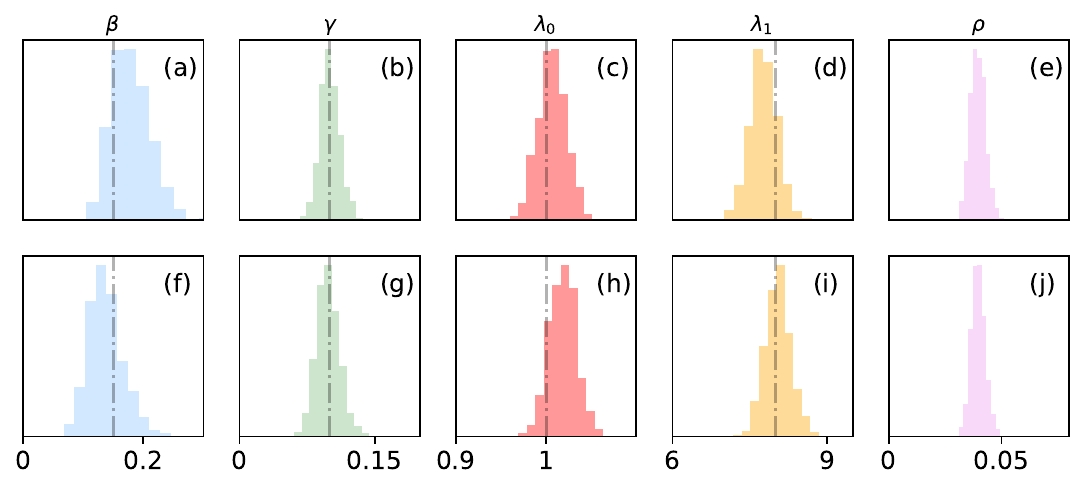}
\caption{NA-MDN-ABC results for two different degree distribution scenarios, for the contagion parameters ($\beta$ and $\gamma$) and the network observation parameters ($\lambda_0$, $\lambda_1$, $\rho$). Panels a)-e) display the results for the Erdős–Rényi network, and Panels f)-j) show results for the log-normal distributed network. Parameters that were utilized in forward-simulation have the ``true" value marked in the histograms with a vertical line.} \label{simulation_histograms}
\end{center}
\end{figure}

In Figure \ref{simulation_histograms}, we plot the marginal posterior densities of disease parameters $\beta$ and $\gamma$, as well as the network parameters $\lambda_0$ and $\lambda_1$, for both network scenarios (Erdős–Rényi and log-normal degree distributions).

\begin{figure}[H]
\begin{center}
\includegraphics[page=1, width=5in]{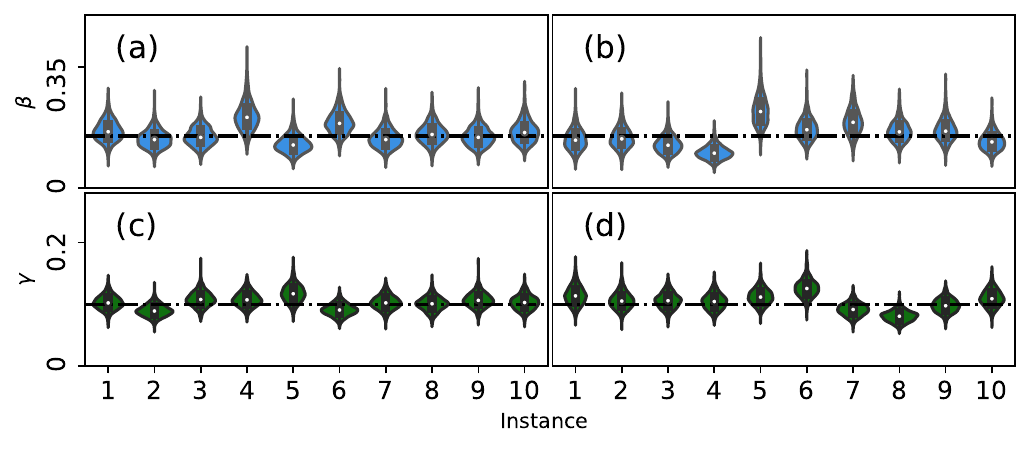}
\caption{NA-MDN-ABC results across 10 instances of the original stochastic epidemic for $\beta$ and $\gamma$ on a Erdős–Rényi network (Panels a) and c), respectively) and for $\beta$ and $\gamma$ on a log-normal network (Panels b) and d), respctively). True values for parameters are marked with a horizontal line.} \label{violin_plots}
\end{center}
\end{figure}

The posterior distributions shown in Figure \ref{simulation_histograms} are much less diffuse than the prior distributions used for the parameters, and tend to be roughly centered around the true values used in the simulation. However, note that the posterior distribution depends on the ``original" realization of simulated epidemic that we use as ground truth. Because this simulation is stochastic in nature, it is possible for the posterior distribution to look different for varying instances of the original epidemic. To visualize this variance, we re-simulate the original epidemic 10 times for each scenario, and re-draw samples using the contact NA-MDN-ABC. We show these distributions in Figure \ref{violin_plots}. 

\begin{figure}[H]
\begin{center}
\includegraphics[page=1, width=2.5in]{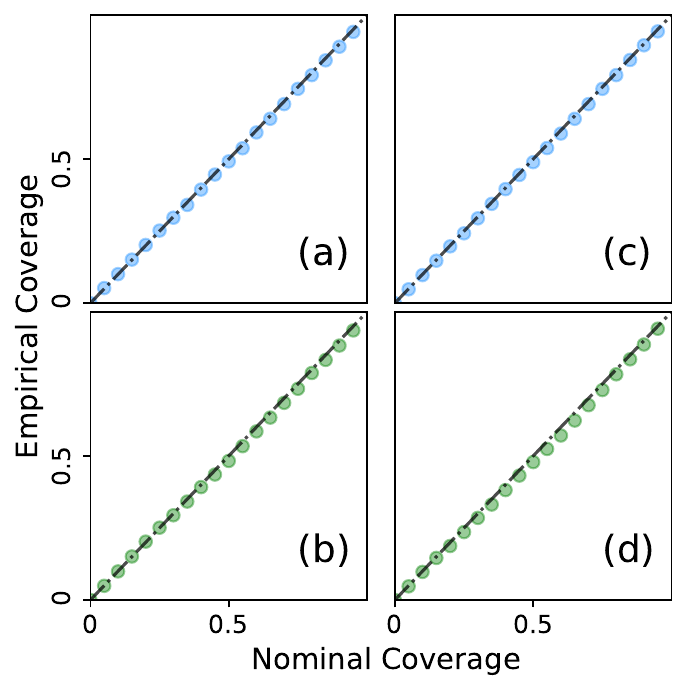}
\caption{Nominal vs. empirical coverages of the MDN-ABC step for a) $\beta$ and c) $\gamma$ for the Erdős–Rényi network, and b) $\beta$ and d) $\gamma$ for the log-normal network. } \label{coverages}
\end{center}
\end{figure}

While the HMC step in NA-MDN-ABC provides exact inferences, the MDN-ABC step is an only an approximation. In order to validate the performance of the MDN-ABC step in the NA-MDN-ABC procedure, we employed a coverage test similar to those proposed in \citep{cook2006, prangle2014_diagnose, talts2018}. 
For 5000 simulation instances, we drew $\theta_0$ from $p(\theta)$ and $(A_{t0},\phi_0)$ from $p(A_t,\phi|X)$. We then simulated a new dataset $Y_0$ based on these parameters, and used these instances of $Y_0$ to draw 5000 posterior samples of $P(\theta,A_t,\phi|X,Y_0)$ via NA-MDN-ABC. We then extracted the $\alpha\%$ credible intervals from these sampled posterior distributions, where $\alpha \in \{0,5,...,95\}$. As defined in \citep{prangle2014_diagnose}, the ``coverage property" is fulfilled for $\theta$ if the ``true" values $\theta_0$ fall in the nominal $\alpha\%$ credible intervals $\alpha\%$ of the time (i.e. the empirical coverage of the $\alpha\%$ credible interval is $\alpha\%$). In Figure \ref{coverages}, we compare the nominal coverage to the empirical coverage for contagion parameters $\beta$ and $\gamma$.

\section{Application: Shark Bay Dolphins}
In this section, we will apply out method to investigate the ongoing transmission of tattoo-skin disease (TSD) in a population of bottlenose dolphins in Shark Bay, Australia. This example is meant to be a demonstration of the NA-MDN-ABC method, rather than a fully accurate investigation of TSD transmission, so some simplifications are made for clarity.

Tattoo-skin disease is an infectious disease that primarily affects cetaceans. It is caused by a variety of cetacean poxviruses \citep{flom1979, bracht2006}, and results in readily visible skin lesions in affected animals \citep{geraci1979}. Previous studies of TSD have linked social behavior to likelihood of being infected by TSD \citep{powell2020}, and TSD prevalence appears to be much higher among nursing calves when compared to weaned juveniles and adults. While zoonotic transmission events have not been observed, humans commonly experience close contact with both wild and captive dolphin populations, making the transmission dynamics of TSD of potential public health interest. Furthermore, dolphins' susceptibility to TSD may be exacerbated by factors such as water temperature and pollution \citep{van2003stress, vanbressem2009}, making TSD prevalence a potential indicator of environmental stressors to marine habitats.

While previous studies have explored the epidemiological traits of TSD, we will apply MDN-ABC to study the individual-level transmission dynamics among the dolphins in Shark Bay, Australia. We will consider the rich dataset collected during the Shark Bay Dolphin Research Project, which tracked sightings of dolphins in Shark Bay starting from 1988, recording dolphin co-proximity events, disease status, and age. A public version of this dataset that contains the co-proximity events and disease statuses can be found at \citep{powell_dataset}. We made use of the data collected between June 1, 2010 and June 1, 2015, excluding observations on dolphins which were not seen in at least 2 sightings, at least 3 months apart. Additional data was kindly provided by the Mann Lab at Georgetown University, which denoted the age category of the dolphins at each sighting: calves (less than 3 years old), juveniles (3 - 10 years old), and adults. The age category of calves, as well as the estimated birthdate of individuals, was based off of sightings of the mothers, size, and ventral speckling \citep{krzyszczyk2012, mcentee2023}. This dataset is fairly unique in its capture of both a contact network of individuals and the individual-level propagation of an ongoing contagion. As with most observational data, it lacks exact event times (i.e. it is unknown exactly when dolphins transition between symptomatic and non-symptomatic states), and the reported co-proximity information serves as a proxy for the true infectious contact of interest (here, physical contact with the skin lesions of an affected dolphin). Thus, we must account for the uncertainty in the observations of the epidemic history as well as the contact network, making the dataset a good candidate for analysis using the NA-MDN-ABC method.

The effects of social behavior on susceptibility to TSD have previously been investigated \citep{powell2020}. While it was found that interactions with infected individuals were predictive of higher risk to TSD infection, the focus of the study was not to quantify the biological mechanisms of TSD transmission. In the following sections, we will analyze the spread of TSD using a simple compartmental disease model applied to a contact netwrok between dolphins. Using MDN-ABC, we will draw posterior inferences on the parameters of the contagion model and compare the results to known literature.

\subsection{Contagion Model}

While the exact biological means of transmission are not fully understood for TSD, we will model TSD as a discrete-time SIR process, with each time-step being one week. 

Because it has been observed that TSD is more prevalent among calves and weaned juveniles than in adults \citep{powell2018}, we consider three separate coefficients for transmissibility: $\beta_c$ for the per-contact transmission probability per time step for calves, $\beta_j$ for the per-contact transmission probability per time step for juveniles, and $\beta_a$ for the per-contact transmission probability per time step for adults. Let $\mathcal{N}$ be the set of all individual dolphins. If at time $t$ each node $i\in \mathcal{N}$ has a disease state $W_t^i\in\{S,E,I,R\}$ and age status $C_t^i \in \{\mbox{calf, juvenile, adult}\}$, this can be expressed as:
$$\beta_c = P(W^i_{t + \Delta t} = E \mbox{ via } j|W_t^i = S, W_t^j = I, C_t^i = \mbox{calf}).$$
$$\beta_j = P(W^i_{t + \Delta t} = E \mbox{ via } j|W_t^i = S, W_t^j = I, C_t^i = \mbox{juvenile}).$$
$$\beta_a = P(W^i_{t + \Delta t} = E \mbox{ via } j|W_t^i = S, W_t^j = I, C_t^i = \mbox{adult}).$$
$$\epsilon = P(W^i_{t + \Delta t} = E \mbox{ spontaneous} | W_t^i = S).$$
Here, $\{W_{t+\Delta t}^i = I \mbox{ via } j\}$ is defined as the event that node $i$ is infected (caused to progress into exposed state) by infected neighbor $j$, while the event $\{W_{t+\Delta t}^i = I \mbox{ spontaneous}\}$ occurs when a node ``spontaneously" acquires the contagion independently of any infectious contacts. This spontaneous probability of infection, defined as the ``spark term" $\epsilon$, captures infectious sources that may not be captured by modeled contact structure (such as infection transmitted by a migratory individual who is never observed by researchers). 

Note that some dolphins transition between age categories during the observation period. While more accurate modeling of the age of dolphins may make use of photographic and social evidence from the dolphin sightings, we make some simplifying assumptions based on the available data. We considered weaned juveniles to be part of the ``juvenile" category for the full duration of observation, unless that individual was later sighted as an adult. In those cases, we consider that juvenile as an adult starting from the first ``adult" sighting. Calves are considered to become juveniles three years after their recorded birth-date. If the birth-date is not available, they become juveniles based on whichever event happens first: 1) three years elapse after their first sighting as a first-year calf, or 2) they are spotted as a weaned juvenile. 

The duration of infectiousness is distributed as a Weibull distribution with shape $\gamma_a$ and scale $\gamma_b$. After this period of infectiousness, individuals transition to a permanent ``recovered" state, at which time they are unable to infect others or be infected. Thus, if $U_i$ is the time that individual $i$ would spend in the infectious period if individual $i$ were to be infected, then:
$$U_i \sim \mbox{Weibull}(\gamma_a, \gamma_b).$$

The event times for the transition into each state is unknown for this population. Instead, the status of each dolphin is only given as a binary indicator of disease presence (visible skin lesions) upon each sighting. While false positives are unlikely, there may be false negatives in the sighting of disease (e.g. the diseased portion of the dolphin is not visible), but we will assume in this example that disease status is reliably reported. We also assume that the symptoms of disease coincide with the ``infected" state in our model, such that infection can only occur upon contact with a symptomatic dolphin. Thus, if $Y_t^i$ is defined as the observed status of node $i$ at time $t$:
$$Y_t^i = 1 \mbox{ if } W_t^i = I \mbox{, } Y_t^i = 0 \mbox{ otherwise}.$$

To initiate our contagion, we treat all dolphins who were spotted with symptoms before and after the June 1, 2010 as the initial infected population. Individuals who were symptomatic before June 1, 2010, but never seen with symptoms afterwards were treated as ``recovered" at $t=0$. This is a fairly simplistic approach to define the initial state of our contagion, as it precludes the possibility that certain individuals may be infected at the start date, but were not spotted with symptoms until afterwards, as well as individuals who may have recovered at $t>0$, but were not seen with symptoms afterwards. Extension of our work for inferring on the starting state of our population is straightforward. For example, the state of individuals (susceptible, infected, or recovered) at initialization can be treated as a latent variable that is then included in the contagion parameters $\theta$. As this example is meant to demonstrate the usage of NA-MDN-ABC rather than serving as a fully-accurate model of the spread of TSD, we will not implement this extension here.

For priors, we set $\beta_c \sim \mbox{Uniform}(0,0.50)$, $\beta_j \sim \mbox{Uniform}(0,0.01)$, $\beta_a \sim \mbox{Uniform}(0,0.002)$, $\epsilon \sim \mbox{Uniform}(0,0.0004)$, $\gamma_k \sim \mbox{Uniform}(0.2,5)$, and $\gamma_\lambda \sim \mbox{Uniform}(10,160)$.

\subsection{Network Observation Model}
In the bottlenose dolphin population of Shark Bay, co-proximity is known to be correlated closely with skin-contact, such that pairs of dolphins spotted together in co-sighting events are also much more likely to be observed engaging in direct contact such as rubbing and playing \citep{leu2020}. However, co-sightings alone (the dataset that is publicly available) cannot be directly used to describe the transmission of TSD, as instances of co-proximity do not capture the potential transmission events for skin disease. Thus, we will utilize the co-proximity events between dolphins to sample from the posterior distribution of the true underlying contact network.

Similar to the example given in Section 3, we will assume that the underlying contact network is unweighted and undirected. Due to potential changes over time in the contact network of dolphins, we will infer a distinct contact network for each year of observation, for a total of five networks. We will consider the observed data to be the number of times each pair of dolphins was observed in a co-proximity event, such that $X_{ijw}$ is the number of times dolphins $i$ and $j$ were spotted in the same group during year $w$. For each year, we model the number of counts as a Negative-Binomial distribution, with the parameters of the distribution varying across the five time-steps. Thus, our network model can be expressed as:
$$X_{ijw} | A_{ijw} = 0 \sim \mbox{NegBin}(n_{0w}, p_{0w}),$$
$$X_{ijw} | A_{ijw} = 1 \sim \mbox{NegBin}(n_{1w}, p_{1w}),$$
$$A_{ijw} \sim \mbox{Bernoulli}(\rho_w).$$
We define network observation parameters $\phi = (n_{01}, ..., n_{05}, p_{01}, ..., p_{05}, n_{11}, ..., n_{15}, p_{11}, ..., p_{15}, \rho_1, ..., \rho_5)$. In Step 2 of Algorithm 1, we draw from the joint posterior of $P(A_{1}, ..., A_{5}, \phi|X)$. This is accomplished using Hamiltonian Monte Carlo, implemeted in STAN, following the method given by \citep{young2020}. In order to break the symmetry between edges and non-edges, and to improve the mixing properties of the algorithm, we will fix $p_{0w} = 1$ for all $w$. Our priors are defined as $n_{0w} \sim \mbox{Gamma}(2,4)$, $n_{1w} \sim \mbox{Gamma}(2,4)$, $p_{0w} \sim \mbox{Gamma}(2,4)$, and $\rho_w \sim \mbox{Beta}(1,20)$ for all $w$.

\begin{figure}[H]
\begin{center}
\includegraphics[page=1, width=5in]{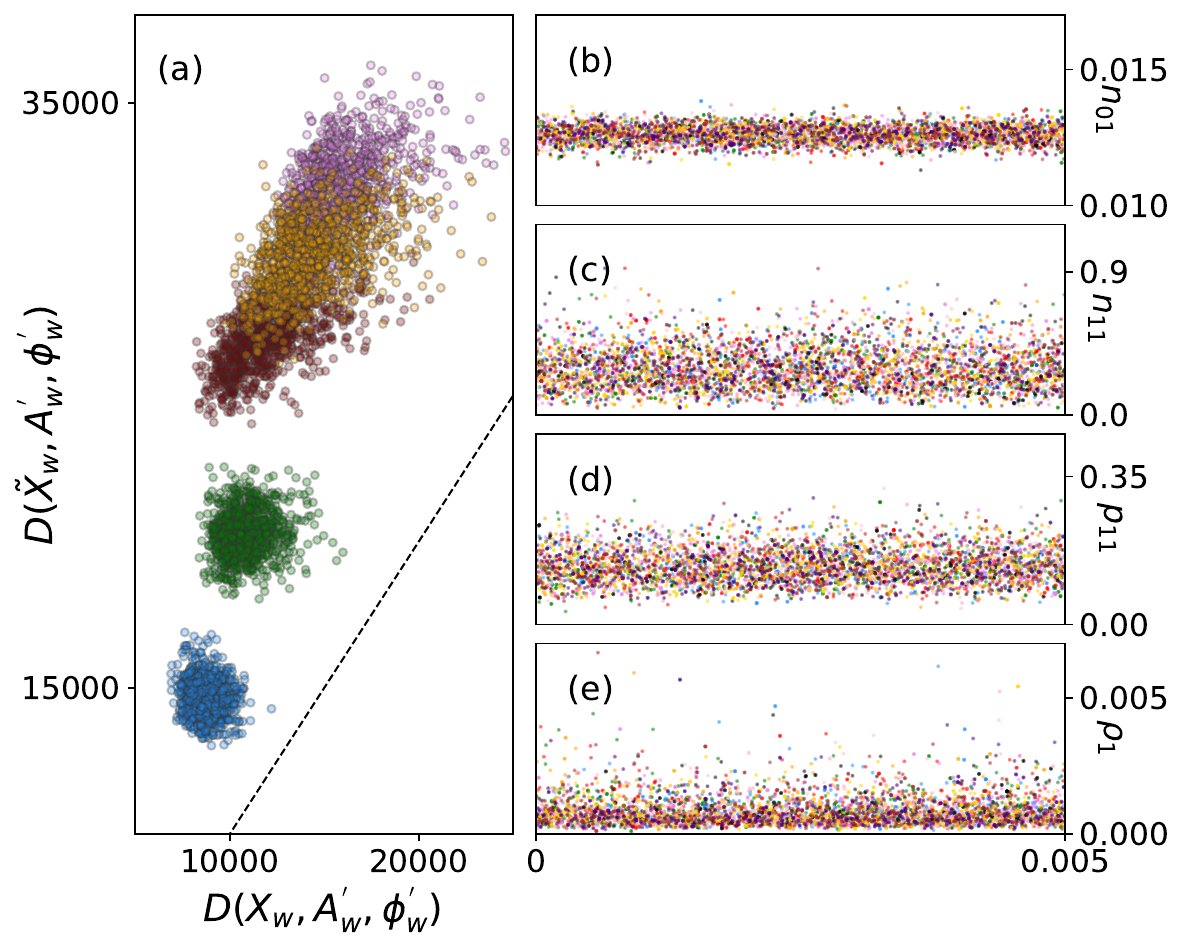}
\caption{In Panel a), discrepancy to the observed data is plotted against discrepancy to data generated from the model, with the dotted line denoting equality. On the right, samples from 10 independent HMC chains are plotted for b) $n_{01}$, c) $n_{11}$, d) $p_{11}$, and e) $\rho$, for time $w = 0$. Note that $p_0$ is set to $p_0 = 1$.} \label{discrepancies}
\end{center}
\end{figure}

In order to validate the choice of model for the network reporting model, we employ the method set out in \citep{young2020} and first proposed in \citep{gelman1996}. Following \citep{young2020}, ``discrepancy" is defined as:
$$D(X,A_{w}^{'},\phi_w^{'}) = \sum_{ij} X_{ijw} \mbox{log}\frac{X_{ijw}}{\langle\Tilde{X}_{ijw}(A_{w}^{'}, \phi_w^{'})\rangle}.$$
Here, $\langle\Tilde{X}_{ijw}(A_{w}^{'}, \phi_w^{'})\rangle$ is the mean of $X_{ijw}$ when the data is generated from the proposed model with the underlying true network set to $A_{w}^{'}$ and the network reporting parameters set to $\phi_w^{'}$. Since a smaller discrepancy implies a better model fit, we draw values of samples of $A_{w}^{'}$ and $\phi^{'}$ from the posterior distribution and calculate the discrepancy between the  $D(X_w,A_{w}^{'},\phi^{'})$ and $D(\Tilde{X_w},A_{w}^{'},\phi_w^{'})$, where $\Tilde{X_w}$ is a network observation matrix simulated from the data-generating model. If the model is a good fit, we would expect $D(X_w,A_{w}^{'},\phi_w^{'})$ to be less than $D(\Tilde{X_w},A_{w}^{'},\phi_w^{'})$ \citep{young2020, gelman1996}. In Figure \ref{discrepancies}, we plot the discrepancies for $w = 1,...,5$, varying colors between the discrepancies calculated for each aggregated duration of time.

Besides checking the goodness of fit for a model, it is also crucial to evaluate the quality of the Markov Chain samples. In Panels b) - e) of Figure \ref{discrepancies}, we plot the samples from the HMC chain, for 10 chains, for the parameters $\phi^{'}_1 = (n_{01}, n_{11}, p_{11}, \rho_1)$. The burn-in period was set to 1000 samples, after which 1000 samples were generated for the plots. Multiple modes or temporal trends may indicate poor mixing or insufficient burn-in, but such signs are absent. Similar results were observed for $w \in \{2,3,4,5\}$.

\subsection{MDN-ABC Settings and Results}

Similar to Section 2, we used a simple feed-forward neural network to learn our summary statistics. We set the dimension of the summary statistics to 20, and trained for 5 gamma-distributed components for our MDN. Once again, we generated 5 million samples for training, and 2.5 million samples for validation. We employed the same training strategy as Section 3.

\begin{figure}[H]
\begin{center}
\includegraphics[page=1, width=4in]{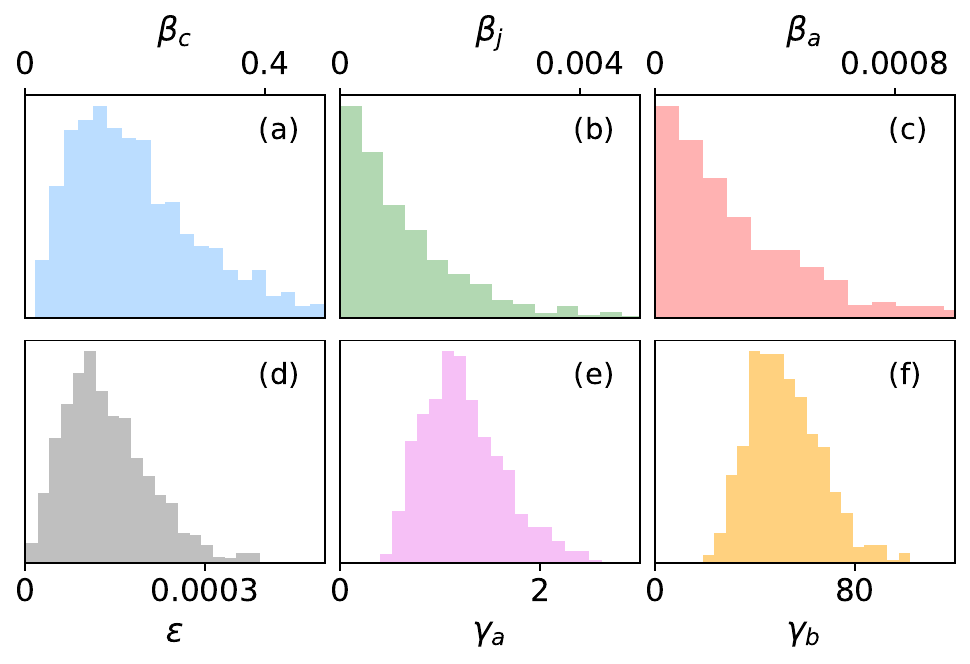}
\caption{NA-MDN-ABC approximate posterior densities for contagion parameters a) $\beta_c$, b) $\beta_j$, c) $\beta_a$, d) $\epsilon$, e) $\gamma_a$, f) $\gamma_b$.} \label{dolphin_histograms}
\end{center}
\end{figure}

To draw the NA-MDN-ABC samples, we re-used the training data and evaluated the summary statistics for each simulation instance. We then calculated the Euclidean distance between each set of summary statistics and the summary statistics yielded by the observed data, and selected the best $0.02\%$ of samples, resulting in approximately 1000 posterior samples. In Figure \ref{dolphin_histograms}, we show the NA-MDN-ABC posteriors for the contagion parameters. Notably the posterior median for $\beta_c$ is nearly 10 times higher than the posterior median for $\beta_j$ and nearly 100 times higher than $\beta_a$. This indicates that calves are far more susceptible to TSD transmission than other age groups, and generally agrees with the observation that the prevalence of TSD is low among adults. In addition, the low values of $\epsilon$ indicate that most of the transmission events can be modeled by transmission via the contact network.

\subsection{Posterior predictive checks and comparison with literature}

\begin{figure}[H]
\begin{center}
\includegraphics[page=1, width=3.5in]{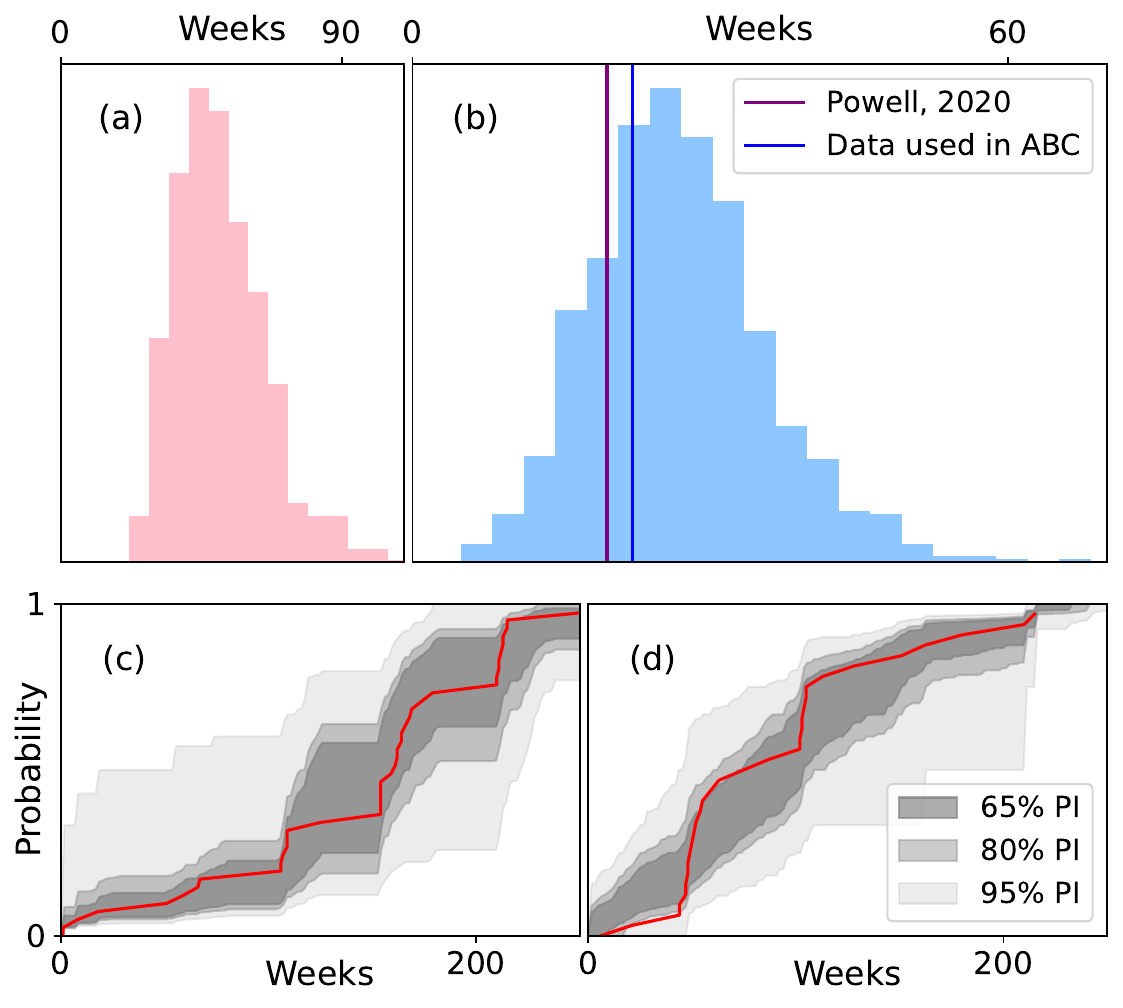}
\caption{Panel a): NA-MDN-ABC approximate posterior density for infectious period. Panel b): Posterior predictive distribution of the mean difference between first and last infectious sightings, compared to \citep{powell2018}. Panel c): Predictive intervals for cumulative distribution function of initial infection times. Panel d): Predictive intervals for cumulative distribution function of interval between first and last infected sightings, for dolphins who were spotted with symptoms at least twice.} \label{dolphin_recovery}
\end{center}
\end{figure}

To the best of our knowledge, this is the first network-based analysis of the individual-level transmission dynamics of TSD among dolphins. Thus, there is no standard of comparison for our inferences on $\beta_c$, $\beta_j$, and $\beta_a$. However, previous literature has considered the infectious period of TSD. In bottlenose dolphins in the Sado estuary, Portugal, it has been estimated that the infectious period lasts between 3 and 45.5 months \citep{vanbressem2003}, though these long symptomatic periods may be due to various sources of pollution in the local environment. In \citep{powell2018}, which studied the Shark Bay bottlenose dolphins, it was estimated from photographic data that the infectious period lasts, on average, 19.6 weeks. By taking the joint distribution of $(\gamma_a, \gamma_b)$ and calculating the posterior distribution of the mean infectious period from our NA-MDN-ABC posterior, we find that the posterior median for the mean infectious period is approximately 51.9 weeks, which appears initially to be a mismatch to the established literature. The posterior distribution of the mean infectious period is shown in the left panel of Figure \ref{dolphin_recovery}.

However, the infectious period in \citep{powell2018} was defined as the difference in time between the first and last times the dolphins were spotted with symptoms of TSD. This is in contrast to our inferences, which aim to capture the full duration of infectiousness, regardless of when the infected dolphins were spotted. Furthermore, while \citep{powell2018} focused on the same Shark Bay population we have analyzed, the authors employed a stricter set of inclusion criteria than we have used here (focusing analysis on a pool of 10 well-observed individuals), as well as a photographic dataset not currently available publicly. In order to compare our model properly with these previous results, we generated 1000 posterior predictive samples by sampling values of $\theta$ and $A_t$ from our MDN-ABC posterior, and used these samples to simulate new epidemic instances $Y_s$ from our model. From these simulations $Y_s$, we calculated the posterior predictive distribution of the mean difference between the first and last times that infected dolphins were spotted with symptoms of TSD (the same method employed above). Furthermore, because the conclusions in \citep{powell2018} stemmed from a different dataset from ours, we performed additional posterior predictive checks to compare the fit of our model to the publicly available observed data used in our analysis. We used the observed data used for our NA-MDN-ABC and once again calculated the mean difference between the first and last times that infected dolphins are spotted with TSD symptoms. This posterior predictive check is shown in Figure \ref{dolphin_recovery}, where the histogram denotes the posterior predictive distribution of this metric. Two vertical lines denote the mean ``observed" symptomatic period derived from both \citep{powell2018} and the data we utilized.

Finally, we observed the posterior predictive properties of two additional metrics. Due to missing event times, we could not directly obtain the ``true" epidemic curve for TSD in this population based on the observed data. Instead, we calculated the distribution of two related metrics: 1) the first time each infected dolphin is initially spotted with symptoms, and 2) the time interval between the first and last times each infected dolphin is spotted with symptoms. Note that this second metric is closely related to the ``mean observed symptomatic period" previously discussed; the difference is that we now consider the entire distribution, rather than the mean only. For both of these metrics, we calculated the cumulative distribution function (CDF) on the observed data, as well as for each of our posterior predictive samples. In Figure \ref{dolphin_recovery}, we can obtain pointwise $\alpha\%$ predictive intervals for the CDFs derived from posterior predictive samples for $\alpha \in \{ 65,80,95 \}$, and evaluate how well these intervals describe the ground truth, plotted in red. In a well-fitted model, it would be expected that the posterior predictive intervals for these metrics would be able to capture the distribution of these metrics calculated from the observed data.

\section{Discussion}
This paper considers the application of a network-augmented MDN-ABC for approximate Bayesian inference for epidemics spreading on noisily observed contact networks. Due to the flexibility of the MDN-ABC and the network augmentation, this method provides a large degree of flexibility in modeling both the spread of contagion and the observation process on the network. While inferences from NA-MDN-ABC are approximate, NA-MDN-ABC allows for posterior inference under complex models, such that researchers are not forced to choose between model expressiveness and analytical tractability.

 There are various directions in which this work can be expanded. First, this paper is primarily concerned with settings where the frequency of disease testing is independent of the underlying contagion parameters. For example, in the Shark Bay Dolphin dataset, the researchers did not choose to specifically target diseased dolphins for observation. However, in some real-world settings, such as during the COVID-19 pandemic, individuals may seek out more frequent testing when they are symptomatic. In such cases, the number of observations (i.e. tests) in $Y$ would vary with values of parameters $\theta$, which would lead to the MDN having an input of varying dimensionality. Our current neural network architecture would be unable to handle this type of complexity, though this could be made possible with more involved architectures, such as recurrent neural networks \citep{mesnil2013}. 
 
Our work primarily considers relatively simple network models. First, the sampling of the unknown network is constrained by the same types of assumptions made in \citep{young2020}; namely, the observations on any dyad are independent of the true status and observed status of any other dyads. While the data model could theoretically be expanded to incorporate such dependencies, this would be computationally expensive. In addition, while we utilized the NA-MDN-ABC for a temporal network in Section 4, our model implicitly assumes that the network is memoryless; the previous relationship between two dolphins does not affect the probability of observing either an edge or a non-edge in future networks. More sophisticated models could incorporate previous network information to gain greater certainty about future networks. 

In the case of the Shark Bay Dolphin dataset, it may also be interesting to consider more advanced models for the spread of TSD. For example, we modeled TSD as a simple SIR disease, which assumes that the incubation period for TSD is short relative to our simulation time-steps (1 week). However, it may be more appropriate to model TSD as an SEIR disease, with an asymptomatic, yet infectious, exposed ``E" state as well. Furthermore, our model does not consider the possibility of different strains of TSD with varying infectivity spreading in the population, though this can be easily implemented. Other extensions may also consider the effects of environmental factors, such as water temperature, on both the transmission and recovery dynamics of TSD. Lastly, we do not consider the possibility of erroneous reporting, where the disease status of dolphins is not known with certainty. While it is unlikely that ``false positives" are observed (i.e. reports of diseased dolphins are reliable), it is possible that some diseased dolphins are reported to be healthy \citep{powell2020}. In particular, because reports on the symptoms of TSD are limited to visual identification of skin lesions, it may be difficult to correct ascertain disease status if an individual's skin lesions are only present on body parts difficult to observe, such as the stomach. Due to the flexibility of NA-MDN-ABC, it is relatively straightforward to implement these types of extensions, as long as these additional model features can be parameterized and forward-simulated.

Lastly, we utilized a simple rejection ABC algorithm for the sake of demonstration. More specialized applications of NA-MDN-ABC may choose to employ more sophisticated algorithms, including Markov Chain Monte Carlo ABC (MCMC-ABC) and Sequential Monte Carlo (SMC-ABC) \citep{marjoram2003, sisson2007}. The primary improvements these algorithms offer is the ability to use previously accepted proposals to inform future proposals, instead of directly sampling from the prior, thus requiring fewer simulations. Because NA-MDN-ABC requires a large set of training simulations to train the MDN, it is natural to simply re-use these training simulations for rejection ABC. However, once the MDN is trained and the summary statistics are obtained, implementation of other ABC algorithms is straightforward.

Every epidemic is unique. The temporal and spatial granularity at which cases are observed, the degree of knowledge about individual contact patterns, and the coverage of testing are all dependent on political and social dimensions of the affected population and biological characteristics of the contagion. In the study of epidemics on networks, there remains a need for methods that are able to bridge the gap between theory and data. To this purpose, we have proposed NA-MDN-ABC as a flexible tool for Bayesian inference on contagion parameters. By delegating the definition of summary statistics to a neural network, this method can be readily extended to a variety of data observation settings to study and model real-world epidemics. 

\newpage

\section{Acknowledgements}
This work was supported by the National Institutes of Health [T32AI007358, R01 AI138901]. We would also like to thank Dr.\ Victor De Gruttola, Dr.\ Ravi Goyal, Dr.\ Till Hoffmann, and the members of the Onnela Lab for their constructive feedback. In addition, we are grateful to Dr.\ Janet Mann, Dr.\ Vivienne Foroughirad, Dr.\ Sarah Powell, Dr.\ Shweta Bansal, and the other members of the Shark Bay Dolphin Project for kindly providing the age-labelled data for the Shark Bay dolphins, as well as their expertise on this dataset. The collection, curation, and analysis of the Shark Bay Dolphin dataset was supported by the National Science Foundation grants 2314823, 1559380, and 1755229.

\section{Data Statement}
The data utilized in Section 4 was primarily drawn from the public dataset provided at \citep{powell_dataset}. Additional data was kindly provided by the researchers of the Shark Bay Dolphin Project, who we reached out to via publicly available contact information.

\section{Appendix}
\subsection{Derivation of Sampling Scheme}
We consider the sampling of the joint posterior $P(\theta, A_t, \phi|X,Y)$.
\begin{equation}
\begin{split}
P(\theta,A_t,\phi|X,Y) & \propto P(X,Y|\theta,A_t,\phi)P(\theta,A_t,\phi) \\
 & = P(X,Y|\theta,A_t,\phi)P(A_t,\phi)P(\theta) \\
 & = P(X|\theta,A_t,\phi)P(Y|\theta,A_t,\phi)P(A_t,\phi)P(\theta)\\
 & = P(X|A_t,\phi)P(Y|\theta,A_t,\phi)P(A_t,\phi)P(\theta)\\
 & = P(Y|\theta, A_t, \phi)P(X, A_t, \phi)P(\theta)\\
 & \propto P(Y|\theta, A_t, \phi)P(A_t,\phi|X)P(\theta).
\end{split}
\end{equation}

Let $(\theta',Y',A',\phi',X')$ be a single instance of the variables sampled from Algorithm 1, and let $D_{\delta,Y}(Y') = 1$ if $||d(S(Y') - S(Y))||\leq\delta$ and 0 otherwise. We can then express the joint distribution of the accepted values of $(\theta',Y',A',\phi')$ as:
$$\pi(\theta',Y',A',\phi') = P(Y'|\theta',A')P(\theta')P(A',\phi'|X)D_{\delta,Y}(Y').$$
Integrating over $Y'$ to marginalize for the unknown parameters, we obtain:
$$\pi(\theta',A',\phi') = \int P(Y'|\theta',A')P(\theta')P(A',\phi'|X)D_{\delta,Y}(Y') dY'.$$
These form the sampled ABC posterior. Note that in the limiting case where $S(Y) = Y$ and $\delta = 0$, then this simplifies to:
$$\pi(\theta',A',\phi') = P(Y|\theta',A')P(A',\phi'|X)P(\theta')\propto P(\theta',A',\phi'|X,Y).$$

\printbibliography

\end{document}